\documentclass[fleqn,10pt]{wlscirep}
\usepackage[utf8]{inputenc}
\usepackage[T1]{fontenc}
\usepackage{subcaption}
\usepackage{amsmath}
\usepackage{graphicx}

\title{Spin-valley coupled thermoelectric energy converter with strained honeycomb lattices}

\author[1]{Parijat Sengupta}
\author[2,*]{Shaloo Rakheja}
\affil[1]{School of Electrical and Computer Engineering, Purdue University, West Lafayette, IN, 47907, USA}
\affil[2]{Electrical and Computer Engineering Department, New York University, New York, NY, 11201, USA}

\affil[*]{shaloo.rakheja@nyu.edu}


\begin{abstract}
A caloritronic device setup is proposed that harnesses the intrinsic spin-valley locking of two-dimensional honeycomb lattices with graphene-like valleys, for instance, silicene and stanene. Combining first-principles and analytic calculations, we quantitatively show that when sheets of such materials are placed on a ferromagnetic substrate and held between two contacts at different temperatures, an interplay between the electron degrees-of-freedom of charge, spin, and valley arises. A manifestation of this interplay are finite charge, spin, and valley currents. Uniaxial strain that adjusts the buckling height in silicene-type of lattices, in conjunction with an applied electric field, is shown to further modulate the aforementioned currents. We link these calculations to a Seebeck-like thermopower generator and obtain expressions (and means to optimize them) for two spin-valley polarized performance metrics--the thermodynamic efficiency and thermoelectric figure of merit. A closing summary outlines possible enhancements to presented results through the inherent topological order and substrate-induced external Rashba spin-orbit coupling that exists in silicene-type materials.
\end{abstract}
\begin{document}

\flushbottom
\maketitle

\thispagestyle{empty}

% \noindent Please note: Abbreviations should be introduced at the first mention in the main text – no abbreviations lists. Suggested structure of main text (not enforced) is provided below.

\section*{Introduction}

\begin{figure}
\centering
\includegraphics[width=4in]{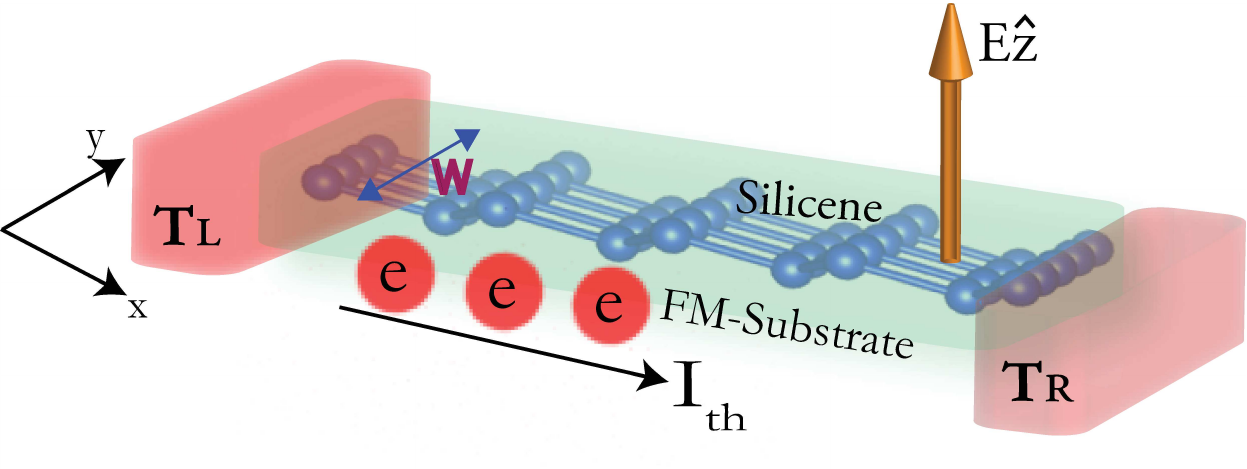}
%\vspace{-2pt}
\caption{The sketch illustrates the suggested arrangement of an X-ene channel flanked between left $\left(L\right)$ and right $\left(R\right)$ contacts maintained at two different temperatures, where $T_{L} > T_{R} $. A thermally-excited electron (hole) current $I_\mathrm{th}$ flows from the contact at a higher (lower) temperature. The channel of length $ L $ and width $ W $ in appropriate units along the \textit{x}- and \textit{y}-axes, respectively is formed by a single layer of silicene, a prototypical X-ene, whose band gap is dynamically tunable via an out-of-plane (\textit{z}-axis directed) electric field $\left(E_{z}\right)$ in conjunction with externally applied uniaxial strain.}
\label{schsi}
%\vspace{-18pt}
\end{figure}

The continuous down-scaling of the feature size of active circuit components necessitates the exploration of optimized cooling pathways to alleviate the problem of localized heating in miniaturized chips~\cite{martinez2015study, topaloglu2019beyond}. The removed heat current can be transformed into useful work taking advantage of the Seebeck effect that constitutes the basis for electric-thermal energy conversion. The optimization of Seebeck-based power and energy conversion techniques,~\cite{cahaya2015spin} of late, have received much attention unifying elements of quantum theory with thermodynamics~\cite{pekola2015towards} as newer materials~\cite{heremans2013thermoelectrics}, notably two-dimensional (2D) graphene, hold promise of better thermoelectric operation demonstrated by a higher thermoelectric figure of merit, \textit{ZT}~\cite{liang2012enhanced,sevinccli2013bottom}. However, the absence of a finite band gap in graphene largely precludes its applicability; this shortcoming is alleviated in other two-dimensional (2D) gapped honeycomb lattices. The laboratory-grown silicene, germanene, and stanene (identified as the X-enes) with a buckled structure, gapped Dirac cones, and graphene-like valleys offer a viable alternative~\cite{ezawa2015monolayer}. In this letter, combining first-principles and analytic calculations, we examine a strained ferromagnetic X-ene based caloritronic device, where the unique spin-valley locking, a hallmark of 2D honeycomb (hexagonal-shaped) lattices, gives rise to a spin- and valley-resolved current. The genesis of such a current $\left(I_\mathrm{th}\right)$ flow lies in a temperature $\left(T\right)$ gradient created difference in the Fermi distribution $\left(f\right)$ at the contacts on either side of a channel. The arrangement is illustrated in Fig.~\ref{schsi} using silicene as the channel material placed on a ferromagnetic substrate. The substrate ferromagnetism, via a Zeeman-type splitting, distinguishes the spin-polarized ensembles located in the vicinity of the $ K $ and $ K^{'} $ valleys. For reasons of a meaningful device realization, we note here that a suitable candidate for the ferromagnetic substrate must be insulating inasmuch as additional leakage paths are cut-off and the thermal current is confined to the X-ene channel only; yttrium-iron-garnet (YIG)~\cite{yang2008magnetic, rao2018thickness} with a high Curie temperature of $ \approx 550\,\mathrm{K} $ and demonstrated ease of integration with 2D material fabrication steps~\cite{leutenantsmeyer2016proximity} can be considered relevant for the present caloritronic setup. Besides, since thermal generation of spin current is envisioned as a key outcome, it is also pertinent to mention about the substrate, spinel ferrite (NZFAO), which supposedly in comparison to YIG-type materials, may improve the transfer of spin across the device~\cite{gray2018spin}.

Briefly, we find a higher spin-down thermoelectric current originating from the $ K^{'} $ valley for two representative X-enes, silicene and stanene. Here, the spin-valley locking~\cite{wang2014multiple,ezawa2013photoinduced} ensures that the corresponding reduced current established on account of carriers located around the $ K $ valley is spin-up polarized. This inequality of two oppositely spin-polarized currents, however, creates a net spin current induced solely via the temperature gradient. We connect these results to the intrinsic Seebeck effect and derive two performance metrics--the thermodynamic efficiency $\left(\varrho\right)$ of the suggested caloritronic setup  and the attendant \textit{ZT}, which are also shown to be valley- and spin-polarized. Additionally, the charge and spin currents along with $ \varrho $ and \textit{ZT} can be amended under the concerted influence of uniaxial strain and a gate-applied electric field. Such dynamic electric field assisted modulations also allow a possible maximization of the caloritronic device performance, the condition for which is succinctly expressed as a ratio between the band gap and the Fermi level. A summary of the key findings and possible enhancements that may include the underlying topological order of X-enes and linked phase transitions appears in the closing section.

\section*{Model and Results}

\begin{figure}
     \centering
     \begin{subfigure}[b]{0.275\textwidth}
         \centering
         \includegraphics[width=\textwidth]{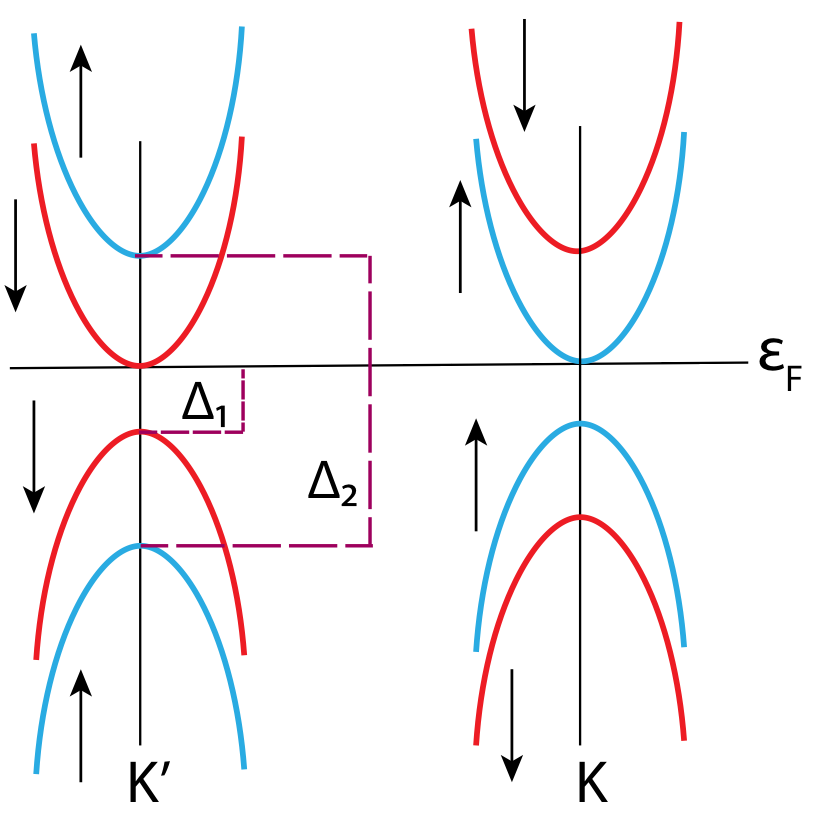}
         \caption{}
         %\label{fig:y equals x}
     \end{subfigure}
   %  \hfill
     \begin{subfigure}[b]{0.35\textwidth}
         \centering
         \includegraphics[width=\textwidth]{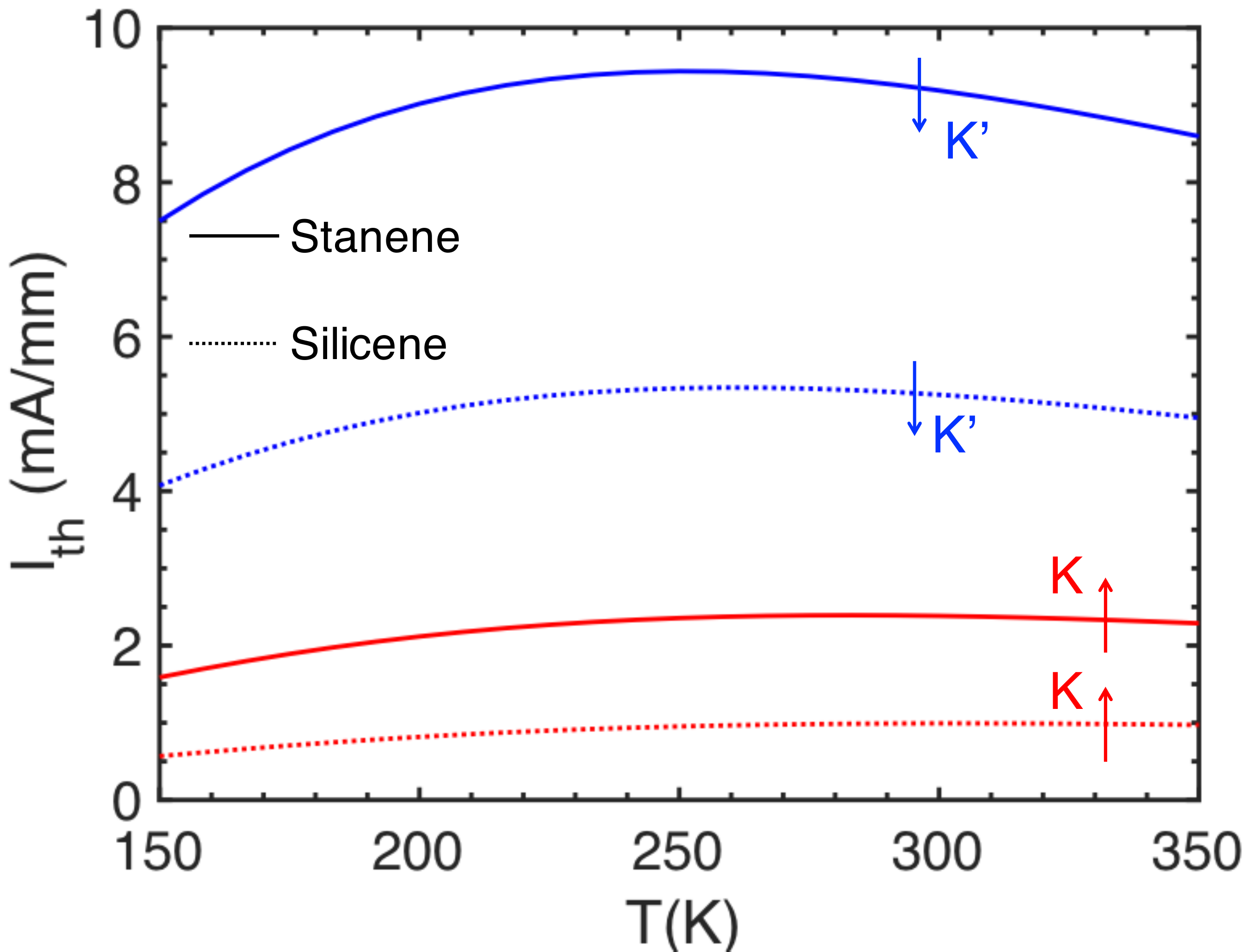}
         \caption{}
        % \label{fig:three sin x}
     \end{subfigure}
    % \hfill
     \begin{subfigure}[b]{0.355\textwidth}
         \centering
         \includegraphics[width=\textwidth]{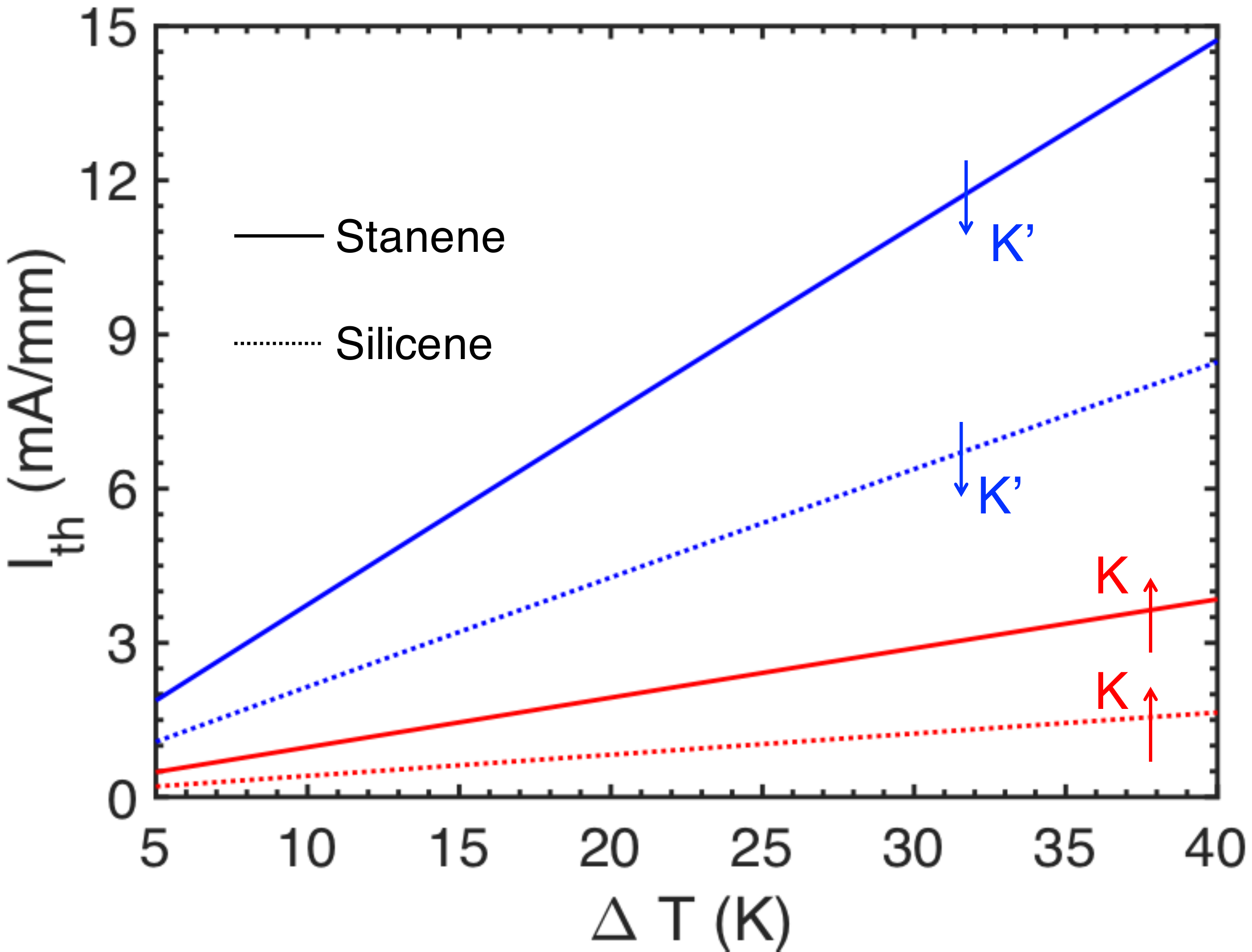}
         \caption{}
         %\label{fig:five over x}
     \end{subfigure}
        \caption{
        (a) shows the spin-polarized nature of the thermal current; the states around the Fermi level $\left(\epsilon_{F}\right) $ carry opposite spins at the $ K $ and $ K^{'} $ valley edges. It is essential to note that unequal spin-defined band gaps ($\Delta_{1} $ and $ \Delta_{2} $) at the valley edges exist on account of the additional $ \ell E_Z \tau_z $ term in Eq.~\ref{ham1} insofar that the spin-polarization ceases for a vanishing $ E_{z} $, the gate field. Such spin-polarization is unattainable for graphene which is strictly planar and exhibits zero buckling. (b) shows the numerically obtained spin and valley resolved thermal current $ I_\mathrm{th} $ that flows in a ferromagnetic X-ene sheet clamped between two contacts at dissimilar temperatures. The temperature $ T_\mathrm{R} $ shown along the \textit{x}-axis is for the right contact; the left contact is always set to $ T_\mathrm{R} + \Delta T $.  
%The necessary material parameters are described in the main text. 
The valley degeneracy is broken by the ferromagnetic exchange energy.
        %whose strength is denoted by $ m\,\mathrm{eV} $ in Eq.~\ref{bgeq}. 
%Lastly, in Fig.~\ref{thcurr}c,
(c)  shows the thermal current behaviour as a function of temperature difference that exists between the two contacts for identical material parameters to those used in (b). The right contact temperature is $ T_\mathrm{R} = 300\, K $ while the left contact temperature is $ T_\mathrm{R} + \Delta T $. A larger temperature gradient $\left(\Delta T\right)$ drives $ \Delta f = f_\mathrm{L} - f_\mathrm{R} $ to a higher value and an increased $I_\mathrm{th}$ for both spin ensembles.}
        \label{thcurr}
\end{figure}

We proceed by writing the model Hamiltonian that represents the setup. The X-enes placed on a ferromagnetic substrate are described by the Hamiltonian~\cite{van2014spin} close to the $ K \left(K^{'}\right) $ valleys by 
\begin{equation}
H=\hbar v_f \left(\eta \tau_x k_x + \tau_y k_y \right) + \Delta \tau_z + m \sigma_z,
\label{ham1}
\end{equation}
where $v_f$ is the Fermi velocity, and $m$ denotes the out-of-plane (z-axis) lattice-ferromagnet exchange field strength. The sub-lattice and spin degree of freedom are denoted by the Pauli matrices $\tau_i$ and $\sigma_i$, respectively. Here $ \eta = 1 (-1) $ identifies the $ K \left(K^{'}\right) $ valley. 
%\textcolor{red}{The ferromagnetic substrate may be implemented using a magnetic insulator, such as yttrium-iron-garnet, which has a high Curie temperature of 550 K and would prevent the shunting of electrons from X-enes due to its insulating nature.}
The strain-tunable band gap is 
\begin{equation}
\Delta = \Delta_\mathrm{so} \sigma_z + m\sigma_z - \ell E_Z \tau_z. 
\label{bgeq}
\end{equation}
Here, $\Delta_\mathrm{so}$ is the intrinsic spin-orbit coupling (\textit{soc}), $\ell$ is the buckling height adjustable via strain, and $E_z$ is the electric field along $z$-axis. The energy dispersion ($\varepsilon-k$) is
\begin{equation} 
\varepsilon = ms \pm \sqrt{\hbar^{2}v_{f}^{2}k^{2} + \Delta^{2}},
\label{disp}
\end{equation}
where $ s = \pm 1 $ and the upper (lower) sign is for spin-up (down) electrons. 
We make use of this dispersion relation to estimate the thermal current flowing in a  strained ferromagnetic X-ene slab clamped between the two contacts (Fig.~\ref{schsi}) maintained at different temperatures ($\Delta T = T_\mathrm{L} - T_\mathrm{R}$). This difference, as alluded to above, creates unequal Fermi distribution in the two contacts and a current $ I_\mathrm{th} $ flows. The subscripts $ L $ and $ R $ indicate the left and right contacts, respectively. An estimate of $I_\mathrm{th}$ in a preset energy ($\varepsilon$) interval is possible by employing the Landauer formalism that relies on the number of transport modes, $ M(\varepsilon)$, and their attendant transmission probability~\cite{datta1997electronic}, $ T\left(\varepsilon\right) $. For an X-ene slab of width $W$, the density of modes (employing Eq.~\ref{disp}) is $ M\left(\varepsilon\right) = \left(\hbar v_f\right)^{-1} \sqrt{\left(\varepsilon \pm m\right)^2-\Delta ^2}$. The lower (upper) sign is for the $ K \left(K^{'}\right) $ valley and $ M_{K}\left(\varepsilon\right) \neq M_{K^{'}}\left(\varepsilon\right) $. 
The transmission probability is set to unity throughout. Assembling these two pieces of information, a numerical estimate for the thermal current, $I_\mathrm{th}$, can be obtained from the following equation:
\begin{equation}
I_\mathrm{th} = \lambda \int M(\varepsilon)d\varepsilon \int_{-\pi/2}^{\pi/2} d\theta \cos(\theta) \left[f_\mathrm{L}(T_\mathrm{L})-f_\mathrm{R}(T_\mathrm{R})\right].
\label{cureq}
\end{equation} 
For brevity, $ \lambda = (eW/\pi h) $; additionally, $\theta$ is the azimuthal angle such that it satisfies the wave vector ($k$) boundary condition: $-k_f \leq k_y \leq k_f$, and $k_y = k\sin(\theta)$. The Fermi wave vector is $k_f$ for a preset Fermi energy ($\varepsilon_F$). We now remark on the form of the current expression; to begin, notice that the presence of the valley-dependent $ M(\varepsilon) $ in Eq.~\ref{cureq} ensures an unequal flow of current such that  $ I_\mathrm{th}^{K} \neq I_\mathrm{th}^{K^{'}} $. This inequality is, however, not limited to the flow of valley-indexed temperature gradient induced charge current--the ramification also extends to the corresponding spin current. To explain the spin-based inequality, recall that the $ K $ and $ K^{'} $ valleys are time reversed pairs, which necessitates that they carry opposite spin parities, from which it follows that the thermally-induced currents originating at the edges would be spin-polarized. The symmetry mandated spin-polarization when combined with the earlier result on unequal valley charge currents leads to the more complete relation, $ I_\mathrm{th}^{K,\uparrow} \neq I_\mathrm{th}^{K^{'},\downarrow} $; this inequality then being the operating principle for the presented caloritronic device (Fig.~\ref{schsi}) that drives a spin- and valley-polarized current. In its entirety, the current flow stands as an explication of an inter-play between the three electron degrees-of-freedom--charge, spin, and valley. Note that the spin current is by definition, $ I_{s} = I_\mathrm{th}^{K,\uparrow} - I_\mathrm{th}^{K^{'},\downarrow} $ and is non-vanishing for the current situation. The spin and valley separable currents rooted in the spin-valley locking of 2D honeycomb lattices is fully embodied in Fig.~\ref{thcurr} obtained via a numerical evaluation of Eq.~\ref{cureq}. Before we analyze the curves in Fig.~\ref{thcurr}, a caveat is necessary: The spin-polarized description of various current types around the valley edges are an outcome of the out-of-plane electric field; to see this note that the arrangement of bands is such that the spin-up band gap (energy difference between the spin-up conduction and valence bands) at $ K $ is identical to the corresponding separation between the spin-down bands at $ K^{'} $. The spin-up and spin-down band gap at $ K $ and $ K^{'} $, respectively is $ \Delta_{1} = 2\vert\left(\Delta_{so} - \ell E_{z}\right)\vert $. The larger band gap (for spin-down at $ K $ and spin-up at $ K^{'} $) not considered is easily computed to be $ \Delta_{2} = 2\vert\left(\Delta_{so} + \ell E_{z}\right)\vert $. The schematic (left sub-figure) in Fig.~\ref{thcurr} with the two band gaps, $ \Delta_{1} $ and $ \Delta_{2} $, sketched clarifies this aspect. For an electric field that closes the band gap~\cite{ezawa2012valley}, this spin-based delineation however disappears and the linked spin-valley polarization ceases. 

The calculations are shown for two X-enes, silicene and stanene, which we assume are uniaxially strained at $ 1.06\% $ and placed on a ferromagentic substrate. The buckling heights from VASP~\cite{hafner2008ab} simulations using the PBEsol exchange functional, a $ 15 \times 15 \times 1 $ $ k $-space mesh, and an energy cutoff of $ 380.0\,\mathrm{eV} $ for silicene and stanene were $ 0.287$ \AA and $ 0.730$ \AA, respectively. The \textit{z}-axis directed electric field applied via a gate electrode is $ E_{z} = 10.0$ mV/\AA. The ferromagnetic exchange energy is $ m = 25.0$ meV and the Fermi level was set to $ \varepsilon_{F} = 80.0$ meV. For a constant $\Delta T = 25$ K between the contacts and an energy window in the range of $ 100 \leqslant \varepsilon\leqslant 200$ meV, the current $I_\mathrm{th}$ as shown in Fig.~\ref{thcurr}(b) is evidently spin and valley polarized. Elaborating further, a higher spin-down current flows around the $ K^{'} $ valley for both silicene and stanene, which is simply an outcome of the greater number of modes available for electrons located there to access in the clamped channel. In addition, notice that the current indicates a rise for low operating temperatures, but at higher values of $ T_\mathrm{R} $ a plateau-like saturation is observed. This behavior is attributable to the diminishing Fermi level difference $\left(\Delta f = f_\mathrm{L} - f_\mathrm{R} \right)$ between the two contacts with a rise in temperature at a given energy in the preset window. While the smearing of the Fermi function is greater at a higher temperature opening up additional modes accessible for transport, the fall in $\Delta f$ (at higher temperatures) executes the more definitive role, essentially pointing to an underplay between $ M\left(\varepsilon\right) $ and $ \Delta f $. Further, exploring another parameter in the design space, the $ I_\mathrm{th}$ profile as anticipated shows an increase in Fig.~\ref{thcurr}(c) in the quantum of current carried by the channel for a larger $\Delta T $, a simple demonstration of an enlarged $\Delta f $ translating into a higher thermal current. 

\begin{figure}
\centering
\includegraphics[width=3.0in]{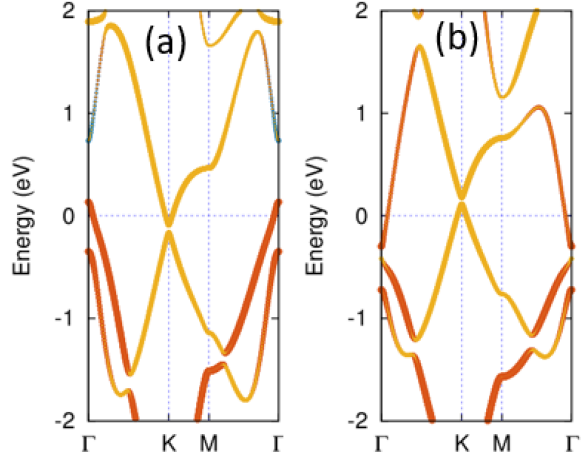}
%\includegraphics[width=3.0in]{trial.pdf}
%\vspace{-3pt}
\caption{The band structure around the K valley for a monolayer stanene under $ 0.98\% $ (a) and $ 1.06\% $ uniaxial strain (b). Stanene with a large $\left(100.0\,\mathrm{meV}\right)$ \textit{soc} induced band gap was chosen to clearly illustrate the shift in dispersion and a topological phase transition. The band gap switches from negative (a) to a positive value (b) through adjustments to the buckling height via strain in conjunction with an electric field.}
\label{bsstan}
%\vspace{-14pt}
\end{figure}

\begin{figure}
\centering
\includegraphics[width=4in]{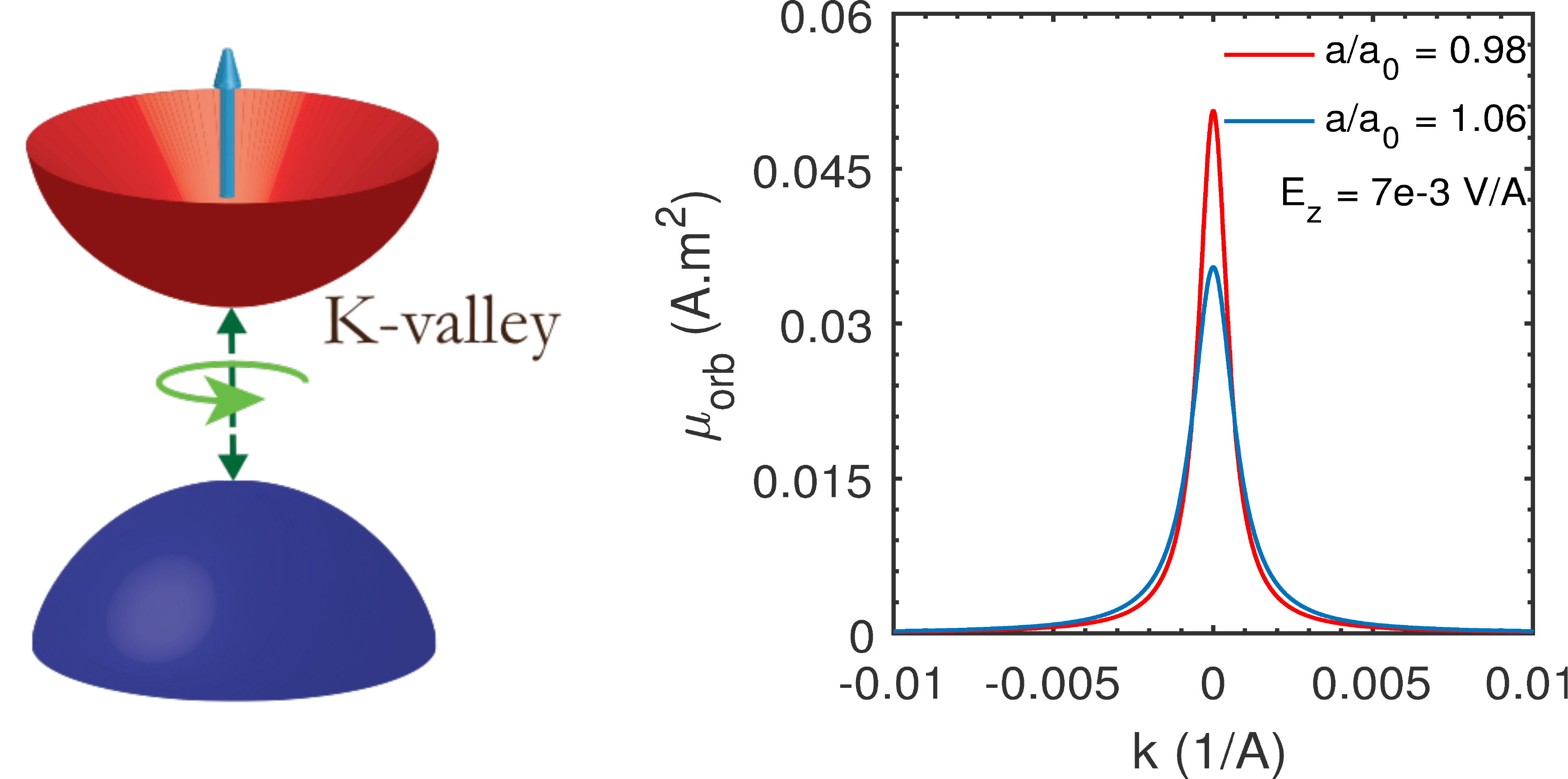}
%\vspace{-3pt}
\caption{The left figure is a sketch of the massive Dirac-like cone formed at the $ K $-valley with a definite spin orientation of electrons - the well-known spin-valley locking. The blue arrow denotes the out-of-plane (\textit{z}-axis aligned) spin which is a good quantum number. The intrinsic magnetic moment $\left(\mathfrak{m}\right)$, which is also pointed out-of-plane is indicated by an anti-clockwise circulation vector. The magnetic moment vector and the spin polarization flips for the time-reversed $ K^{'} $-valley. The $ \mathfrak{m}\left(k\right) $ at the $ K $-valley for a representative X-ene, stanene, using the result of Eq.~\ref{magmom} for two values of strain and an electric field (noted in the plot) is shown in the right figure; at the $ K^{'} $ edge the magnitude remains identical but with a change of sign.}
\label{magfig}
%\vspace{-14pt}
\end{figure}

At this point, a set of comments, supplementary to this discussion is in order: Firstly, a spin-valley polarized current can also be accomplished if the channel in Fig.~\ref{schsi} is substituted by a double-valley hexagonal monolayer of transition metal dicahlcogenides, for example, WSe$_{2} $, where the intrinsic \textit{soc} is a significant contributor. However, a dynamically adjustable current is possible exclusively for X-enes through an amenable density of modes (and dispersion) via an external gate-applied electric field and mechanical strain~\cite{modarresi2015effect}. More interestingly, for suitable combination of strain and electric field, the band gap in an X-ene can toggle between positive and negative values, the switch to the latter leading to possible presence of topological edge states~\cite{tahir2013valley}. While we do not pursue thermoelectrics with X-ene edge states here, however, as an illustration of this topological phase transition, the band dispersion of stanene for two values of uniaxial strain is shown in Fig.~\ref{bsstan}. It is easy to see the band gap change sign (from `-' to `+') at the $ K $ valley edge as the applied strain rises, transforming the X-ene from a topological to normal insulator phase.
The second comment pertains to another `control mechanism' to further amend the spin- and valley-polarized currents. We take advantage of the magnetic moment intrinsic to honeycomb 2D lattices. Briefly, in addition to a spin magnetic moment, the Bloch electrons carry an orbital magnetic moment arising from the self-rotation~\cite{sundaram1999wave,xiao2010berry} of the wave packet, which in case of 2D honeycomb lattices is directed out-of-plane and switches sign at the valley edges, as evident from Eq.~\ref{magmom}. This is given as~\cite{xiao2012coupled}
\begin{equation}
\mathfrak{m}\left(k\right) = \pm\dfrac{e\hbar v_{f}^{2}\Delta}{2\left[\hbar^{2}v_{f}^{2}k^{2} + \Delta^{2}\right]}.
\label{magmom}
\end{equation}
The upper (lower) sign is for the $ K \left(K^{'}\right) $. An out-of-plane magnetic field $\left(\mathbf{B}\right)$ couples to $ \mathfrak{m}\left(k\right) $ to alter the band gap (Eq.~\ref{bgeq}) as $ \Delta^{'} = \Delta \pm \mathfrak{m}B $. A plot of the magnetic moment for stanene calculated using Eq.~\ref{magmom} is shown in Fig.~\ref{magfig}. Clearly, the production of unequal band gaps (it increases at the $ K $ valley) rearranges the dispersion (Eq.~\ref{disp}), creating greater asymmetry in the valley-specific $ M\left(\varepsilon\right) $ and the attendant spin-valley polarized charge currents. 

%\begin{figure}
%     \centering
%     \begin{subfigure}[b]{0.275\textwidth}
%         \centering
%        % \includegraphics[width=\textwidth]{massive.pdf}
%                 \includegraphics[width=1.75in, height=2.25in]{massive.pdf}
%         \caption{}
%         %\label{fig:y equals x}
%     \end{subfigure}
%  %  \hfill
%     \begin{subfigure}[b]{0.35\textwidth}
%         \centering
%         \includegraphics[width=3.3in, height=2.7in]{magnmom.pdf}
%         \caption{}
%         %\label{fig:five over x}
%     \end{subfigure}
%        \caption{The left figure is a sketch of the massive Dirac-like cone formed at the $ K $-valley with a definite spin orientation of electrons - the well-known spin-valley locking. The blue arrow denotes the out-of-plane (\textit{z}-axis aligned) spin which is a good quantum number. The intrinsic magnetic moment $\left(\mathfrak{m}\right)$, which is also pointed out-of-plane is indicated by an anti-clockwise circulation vector. The magnetic moment vector and the spin polarization flips for the time-reversed $ K^{'} $-valley. The $ \mathfrak{m}\left(k\right) $ at the $ K $-valley for a representative X-ene, stanene, using the result of Eq.~\ref{magmom} for two values of strain and an electric field (noted in the plot) is shown in the right figure; at the $ K^{'} $ edge the magnitude remains identical but with a change of sign.}
%        \label{magfig}
%\end{figure}

\section*{Additional thermoelectric performance metrics}
In the preceding paragraphs, a framework was presented to estimate the quantum of current flowing on account of a temperature gradient, which is also a microscopic demonstration of the well-known Seebeck effect. We formally derive the Seebeck coefficient $\left(\mathcal{S}\right)$ for strained X-enes and link it to the thermodynamic efficiency of the proposed caloritronic setup in Fig.~\ref{schsi}. An allied quantity, the thermoelectric figure of merit, \textit{ZT}, often used to gauge thermoelectric performance is also examined. Firstly, $\left(\mathcal{S}\right)$ in the limit, $ k_{B}T \ll \varepsilon_{F} $ is~\cite{lofwander2007impurity} 
\begin{equation}
\mathcal{S} = -\dfrac{\pi^{2}}{3e}\dfrac{k_{B}^{2}T}{\sigma}\dfrac{\partial \sigma}{\partial \varepsilon}.
\label{mott}
\end{equation}
A numerical assessment of $ \mathcal{S} $ must begin with an explicit evaluation of the (intra-band) Drude conductivity $\left(\sigma\right) $ in Eq.~\ref{mott}; we evaluate it with the Kubo conductivity expression~\cite{bruus2004many} for a non-interacting sample in 2D $ k $-space. The generalized relevant Kubo conductivity expression is:
\begin{flalign}
\sigma^{\alpha\beta} =  \gamma\sum_{n,n^{'}}\dfrac{f\left(\varepsilon_{n}\right)- f\left(\varepsilon_{n^{'}}\right)}{\varepsilon_{n} - \varepsilon_{n^{'}}}\dfrac{\langle\,n\vert\,\hat{v}_{\alpha}\vert\,n^{'}\rangle \langle\,n^{'}\vert\,\hat{v}_{\beta}\vert\,n\rangle}{\varepsilon_{n} - \varepsilon_{n^{'}}+i\,\eta},
\label{kubof} 
\end{flalign}
where $\gamma = -i\hbar\,e^{2}/L^{2}$, $ \vert\,n\rangle $ and $ \vert\,n^{'}\rangle $ are eigen functions derived by diagonalizing the Hamiltonian (Eq.~\ref{ham1}), and $ \eta $ represents a finite broadening (lifetime of the electron on the Fermi surface) of the eigen-states resulting from surface imperfections and impurity scattering. For intra-band conductivity, we set $ \vert\,n\rangle = \vert\,n^{'}\rangle$. To clarify choice of notation, the superscripts on $ \sigma $ mean that upon application of an electric field along $ \hat{e}_{\beta} $, the electric conductivity tensor gives the current response along $ \hat{e}_{\alpha} $. We have also tacitly assumed that the wave functions retain their pristine form, the presence of impurities and defects notwithstanding. For a numerical evaluation of the Drude conductivity in the case of X-enes, we must insert the appropriate eigen functions and eigen energies in Eq.~\ref{kubof}. The matrix element, $ M = \langle\,n\vert\,\hat{v}_{x}\vert\,n\rangle $, is therefore $ -\hbar v_{f}^{2}k\sin\theta/\sqrt{\Delta^{2} + \left(\hbar v_{f}k\right)^{2}}$. In obtaining the above expression, we have chosen the conduction band states to evaluate the matrix product; this choice is made by setting the Fermi level to bottom of the conduction band. Inserting the matrix element in Eq.~\ref{kubof}, the Drude conductivity is 
\begin{equation}
\sigma_{D} = \dfrac{\left(e\hbar v_{f}\right)^{2}}{4\pi^{2}\hbar\eta}\int_{0}^{2\pi}\sin^{2}\theta d\theta\int kdk\dfrac{\left(\hbar v_{f}k\right)^{2}}{\Delta^{2} + \left(\hbar v_{f}k\right)^{2}}\dfrac{\partial f}{\partial \varepsilon}.
\label{dr1}
\end{equation}
In Eq.~\ref{dr1}, we have replaced the summation over momentum states by the integral; additionally the term $f\left(\varepsilon_{n}\right)- f\left(\varepsilon_{n^{'}}\right)/\left(\varepsilon_{n} - \varepsilon_{n^{'}}\right) $ is approximated as $ \partial f/\partial \varepsilon = -\delta\left(\varepsilon_{f} - \varepsilon\right) $ by Taylor expanding the Fermi distribution, $ f\left(\varepsilon_{n^{'}}\right) = f\left(\varepsilon_{n} \right) + \left(\varepsilon_{n^{'}} - \varepsilon_{n}\right)\partial f/\partial \varepsilon $. Note that the relation $ \partial f/\partial \varepsilon = -\delta\left(\varepsilon_{f} - \varepsilon\right) $ holds exactly at $ T = 0 $. Converting the $ k $-space integral to energy space using the dispersion relation for an X-ene (with $ m = 0 $) and integrating out the angular variable $ \left(\int_{0}^{2\pi}\sin^{2}\theta\,d\theta = \pi\right)$, we rewrite Eq.~\ref{dr1} (normalized to $ e^{2}/h $ ) as 
\begin{equation}
\sigma_{D} = \dfrac{1}{2\eta}\int d\varepsilon\dfrac{\varepsilon^{2}-\Delta^{2}}{\varepsilon}\delta\left(\varepsilon_{f} - \varepsilon\right) = \dfrac{\Omega}{2\eta},
\label{dr2}
\end{equation}
where $ \Omega = \left[\varepsilon_{f}^{2}-\Delta^{2}\right]/\varepsilon_{f}$. Writing the broadening parameter, $ \eta = \hbar/\tau_{tr} $ reproduces the expression in form of Drude conductivity. Substituting for $ \sigma $ from Eq.~\ref{dr2} in Eq.~\ref{mott} and working out the derivative, the thermopower expression simplifies to
\begin{equation}
\mathcal{S} = -\dfrac{2\pi^{2}}{3e}\dfrac{\eta\,k_{B}^{2}T\varepsilon_{f}}{\varepsilon_{f}^{2}-\Delta^{2}}\dfrac{\partial \sigma}{\partial \varepsilon} = -\dfrac{\pi^{2}}{3e}k_{B}^{2}T\dfrac{1+t}{1-t}\dfrac{\sqrt{t}}{\Delta}.
\label{motttm}
\end{equation}
In Eq.~\ref{motttm}, $ \sqrt{t} = \Delta/\varepsilon_{F} $. We utilize this result to obtain the efficiency of the suggested caloritronic device.
\begin{figure}[t!]
\centering
\includegraphics[width=2.8in]{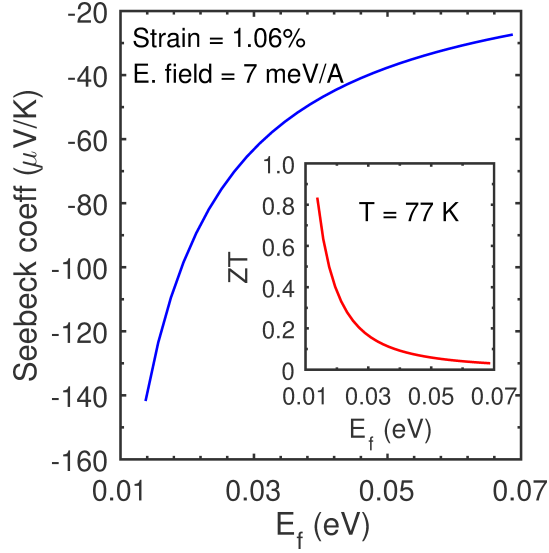} 
%\vspace{-6pt}
\caption{The numerically calculated (using the Mott expression) Seebeck coefficient $\left(\mathcal{S}\right)$ is shown for strain and electric field values indicated on the plot. The inset depicts the corresponding \textit{ZT} value (T = 77 K). Both $ \mathcal{S} $ and \textit{ZT} are dynamically adjustable; for instance, as illustrated here, with a rise in Fermi energy, the S coefficient and the \textit{ZT} drops. Strain engineering in buckled honeycomb lattices with an electric field that tunes the band gap is another possible alternative.}
\label{szt}
%\vspace{-18pt}
\end{figure} 

For a load $ R_{L} $ connected to the device, the transfer of output electric power can be derived by noting that the current is $ I_{c} = \left(W \times L\right)\sigma \mathcal{S}\Delta T/L $, where the cross-sectional area is $ W \times L $ and the electric field is $ \mathcal{S}\Delta T/L $. The total output power is therefore $ P_{op} = \left(\sigma\mathcal{S}W\Delta T\right)^{2}R_{L} $. The input heat power flowing in to maintain the gradient is $ P_{in} = W \kappa \Delta T $. The thermodynamic (Carnot-like) efficiency is:
\begin{equation}
\varrho\left(t\right) = \dfrac{\pi^{4}k_{B}^{4}R_{L}}{18h\eta\mathcal{L}\Delta^{2}}\left[\dfrac{1+t}{1-t}\right]^{2}tT\Delta T\Omega.
\label{theff}
\end{equation}
Here, the ratio $ \kappa/\sigma $ was simplified by taking recourse to the Wiedemann-Franz law (WFL), which is $ \kappa/\sigma = \mathcal{L}T $.~\cite{chester1961law} The thermal conductivity is $ \kappa $ while $ \mathcal{L} = \pi^{2}k_{B}^{2}/3e^{2} $ denotes the Lorentz number. The parameter $ \varrho $ as a function of $  t = \left(\Delta/\varepsilon_{F}\right)^{2} $ can be easily preset and optimized by a suitable tweaking of the band structure. Likewise, the closely linked performance metric, the thermoelectric figure-of-merit, is expressed as $ ZT = \left(\mathcal{S}^2\sigma/\kappa\right)T $; inserting $ \mathcal{S} $ from Eq.~\ref{motttm} and noting that $ \sigma/\kappa = 1/\mathcal{L}T $, we obtain the following expression:
\begin{equation}
ZT = \dfrac{\pi^{2}k_{B}^{2}T^{2}}{3}\left[\dfrac{1+t}{1-t}\right]^{2}\dfrac{t}{\Delta^{2}}.
\label{ztexp}
\end{equation}
We plot (Fig.~\ref{szt}) the Seebeck coefficient $\left(\mathcal{S}\right)$ employing the Mott result The inset shows the attendant $ZT$ of this arrangement. The $ ZT $ drops with an increase in $\varepsilon_F$, a trend that matches the behaviour of $ \vert \mathcal{S} \vert $, which also exhibits a similar flattening out. Further, notice that the   $ \varrho $ and \textit{ZT} expressions have a functional dependence on $ t $ through the term $ t\left(1+t/1-t\right)^{2} $; it may therefore be possible to obtain a condition that maximizes the two performance indicators. We take up \textit{ZT} and evaluating the extremum condition, $ \dfrac{d}{dt}\left(ZT\right) = 0 $, obtain the following equation
\begin{equation}
t^{3} - 3t^{2} - t + 1 = 0.
\label{maxmin}
\end{equation}
The above cubic has three solutions: $ t = -0.675, 0.4608, 3.214 $. Enforcing the requirement $ t = \Delta^{2}/\varepsilon_{F}^{2} > 0 $, which allows us to discard the negative solution, the condition for a maximum $\left(\left(d^{2}/dt^{2}\right)ZT < 0\right) $ is fulfilled for $ t = 0.4608 $. An optimal thermoelectric performance is achievable when the adjustable ratio of $ \sqrt{t} = \Delta/\varepsilon_{F} = 0.678 $. Since both $ \Delta $ and $ \varepsilon_{F} $ can be simultaneously gate-controlled - the former through the buckling height contribution - a dynamic modulation of the proposed caloritronic engine is feasible. 

At this stage, the analysis of the three quantities, namely, $ \mathcal{S} $, $  \varrho $, and \textit{ZT} is complete; we now show that these quantities also can be spin-dependent and consequently valley-locked. To begin, note that they explicitly demonstrate a dependence on the band gap that can be adjusted specific to a valley, a point which we alluded to previously in our discussion (cf. Fig.~\ref{magfig}) of orbital magnetic moments for gapped Dirac materials. It is also tenable to create unequal band gaps at the valley edges through an external driving field, for example, an intense laser beam can rearrange the bands inasmuch as a photon-dressed graphene-like feature of massless Dirac cone ensues. For intermediate strengths of light irradiation, a simultaneous shrinking and amplification of the band gaps at the valley edges can occur, a precise numerical determination of which is possible by resorting to the Floquet formalism.~\cite{kitagawa2011transport,cayssol2013floquet} An illustrative calculation where such photo-aided modulations to the band gaps were carried out with monolayer transition metal dichalcogenides (a hexagonal/honeycomb system) can be found in Ref.~\citenum{sengupta2016photo}. 
\begin{figure}[t!]
\centering
\includegraphics[width=3.5in]{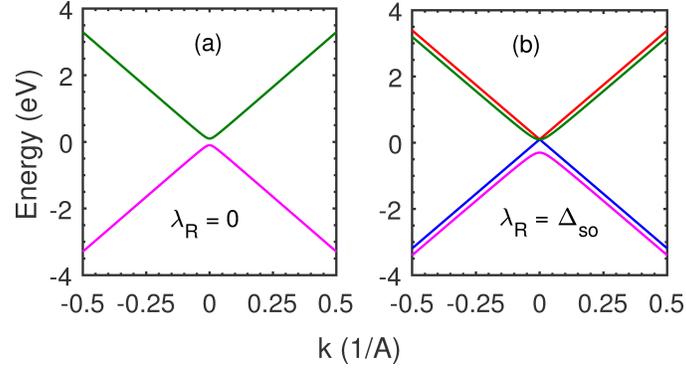} 
%\vspace{-6pt}
\caption{The band structure of stanene (on a substrate) under the combined influence of intrinsic spin-orbit coupling (\textit{soc}) and structural imperfection induced Rashba splitting. The spin-orbit coupling is $ 0.1\,\mathrm{eV} $ for all sub-figures. The Fermi velocity is $ v_{f} = 10^{6}\,m/s $. For vanishing $ \lambda_{R} $ in the left figure (a), a finite gap of $ 2\Delta_{so} $ exists and the spin-up and spin-down bands (conduction and valence) are degenerate. For the right figure (b), the tunable $ \lambda_{R} $ is set to the stanene \textit{soc}; the four bands split and a gap closing emerges indicating topological edge state behaviour. In these plots, the gate field and magnetization has been turned off (cf. Eq.~\ref{bgeq}) and do not contribute to the overall band gap. The Rashba field arises solely from the intrinsic inversion asymmetry between stanene, the representative X-ene, and the substrate.}
\label{topfig}
%\vspace{-18pt}
\end{figure} 

\section*{Discussion}
Summarizing, we have predicted the quantum of thermally-driven spin-resolved current in  strained ferromagnetic X-enes held between contacts at different temperatures. These results not just serve as an illustration of the microscale energy harvesting paradigm (thermal-to-electric) via a prototypical Seebeck-like power generator but also emphasize on the platform offered via the spin-valley locking inherent in X-enes for designing advanced thermal management schemes mediated by the electron spin – the emerging field of spin caloritronics.~\cite{bauer2012spin,boona2014spin} This work can be further expanded through a 1) microscopic optimization of the $ \varrho $ and \textit{ZT} parameters from a wide selection of 2D material systems and substrates (as a case in point, layered black phosphorus~\cite{ling2015renaissance}) probed using first-principles calculations and 2) examination of topological structures that may emerge from the previous study and their thermoelectric applicability. We briefly explain the last point using X-ene-like 2D sheets. Firstly, it is well understood that an X-ene sheet (including graphene) placed on a substrate that breaks mirror and inversion symmetry induces a Rashba interaction, reflected in an additional term in the original Hamiltonian (Eq.~\ref{ham1}) as shown below~\cite{ertler2009electron,pan2014valley}
\begin{equation}
H_{eff} = \hbar\,v_{f}\left(\tau\,k_{x}\sigma_{x} + k_{y}\sigma_{y}\right)\otimes \mathbb{I} + \Delta \tau_z + \lambda_{R}\left(\tau\,\sigma_{x}\otimes s_{y} - \sigma_{y}\otimes s_{x}\right). 
\label{baseqn}
\end{equation}
In Eq.~\ref{baseqn}, $ \mathbb{I} $ is the $  2 \times 2 $ identity matrix, $ \lambda_{R} $ is the Rashba spin-orbit coupling constant and owes its origin to structural inversion asymmetry (SIA), such as charged impurities and adsorbed atoms. The iso-spin matrices are denoted by $ \sigma $ while $ s_{x} $ and $ s_{y} $ represent the Pauli spin matrices. The two sub-lattices are identified by the parameter $ \tau = \pm $. The Fermi velocity v$_{f} $ is assumed to be isotropic. The Hamiltonian is of size $ 4 \times 4 $ in the expanded basis set of iso-spin and spin space $ \left(\vert A\rangle,\vert B\rangle\right)\otimes \left(\uparrow,\downarrow\right) $. The four eigen values, $ E_{j = 1-4} $, of the Hamiltonian in Eq.~\ref{baseqn} which correspond to the spin-up and spin-down components of the conduction and valence bands are
\begin{equation}
E_{j} = s\lambda_{R} \pm \left(\sqrt{\left(\hbar\,v_{f}k\right)^{2} + \left(\lambda_{R} - s\Delta_{so}\right)^{2}} \right). 
\label{eigvalimp}
\end{equation} 
The symbol $ s = \pm 1 $ indicates the spin-up and spin-down states while the upper (lower) sign represents the conduction (valence) bands. We plot two cases (Fig.~\ref{topfig}) using Eq.~\ref{eigvalimp} to underscore the presence of possible gap-closing topological edge states. Easily, for a finite $ \lambda_{R} $ (Fig.~\ref{topfig}b) such that it matches the intrinsic \textit{soc} a zero band gap situation occurs - a probable starting point for further analyses of topological behaviour with more complicated substrates that bring forth large rotational and translational mismatches with the X-enes. Beginning with the parameterization of the conjoined Hamiltonian (substrate and X-ene) from first-principles calculation, insight can be gained into their topological fermion-matter-guided thermoelectrics via a tabulation of \textit{ZT}-like quantities that may display anomalous character. Moreover, complementary to the foregoing predicted theoretical and computational outcomes, from an experimental standpoint, the laboratory measurement of thermoelectric coefficients including those not discussed here, such as the Nernst and Rigi-Leduc effect, offers an indirect probe of topological rearrangements and statistical variations thereof through linked phase transitions.

%\bibliography{sample}

\section*{Author contributions statement}

P.S. and S.R. conceived the device setup and performed the calculations. P.S. and S.R. analysed the results. Both authors reviewed the manuscript. 

\section*{Competing financial interests}
The authors declare no competing financial and non-financial interests.

\end{document}